\documentstyle[preprint,prc,aps,psfig]{revtex}

\renewcommand{\vec}[1]{\mbox{\boldmath $#1$}}

\tightenlines

\begin{document}

\title{Structure of positive energy 
states \\ in a deformed mean-field potential} 
\author{K. Hagino$^{1,2}$ and Nguyen Van Giai$^2$}
\address{$^1$Yukawa Institute for Theoretical Physics, Kyoto
University, Kyoto 606-8502, Japan }
\address{$^2$Institut de Physique Nucl\'eaire, IN2P3-CNRS, \\ 
Universit\'e Paris-Sud, F-91406 Orsay Cedex, France}

\maketitle

\begin{abstract}

We investigate the properties of single-particle resonances in a 
non-spherical potential by solving the coupled-channels equations for the 
radial wave functions. 
We first generalize the box discretization method for positive 
energy states to a deformed system. As in the spherical case, we find 
that the discretized energy is stabilized against the box size when a 
resonance condition is met. 
Using the wave functions thus obtained, we then discuss the energy and the 
radial dependences of scattering wave functions in the vicinity of 
an isolated resonance. In the eigenchannel basis, where the $S$-matrix is 
diagonal, we propose a generalized expression for the 
factorization formula for the multi-channel wave function. 
We find that the factorized wave function agrees well with the exact solution 
inside the centrifugal barrier when 
the energy distance from the resonance is less than the resonance width.
\end{abstract}


\section{INTRODUCTION}

Recent experimental activities on drip-line nuclei have opened up 
a renewed interest in 
the continuum spectra of finite many-body systems. 
These nuclei are often weakly bound, or may even be 
unbound, and a proper treatment of the 
positive-energy states is crucial for their theoretical description. 
The continuum influences the properties of loosely bound nuclei 
mainly through the resonant states, which have a dominant contribution
both in the ground state, e.g., through the correlations induced by the
virtual pair scattering, and in the excited states 
\cite{BLSV02,MNPB02,D96,GSGL01,SGL00,SLW97,BDP99,WD97,ASN97}. 
It is thus becoming more and more important to approach 
the structure of unstable nuclei from the nuclear reaction point of 
view\cite{MW69,OPR03}. 
Recent developments of shell model calculations for weakly-bound systems 
by Id Betan {\it et al.} \cite{BLSV02} and by Michel {\it et al.} 
\cite{MNPB02} clearly indicate this direction. 

A single-particle resonance state can be obtained relatively easily 
when the mean field potential has a spherical shape. In this case, 
the single-particle angular momentum is conserved, 
and the Schr\"odinger 
equation is reduced to a one dimensional differential equation 
for the radial motion. Approximated resonance states are easily found 
by putting a system in a box and imposing a condition that the wave 
function vanishes at the edge of the box. With this boundary condition, 
the continuum spectra are discretized. It has been recognized that, 
for a resonance state, the single-particle 
energy is stabilized against the box size, while 
it changes significantly as a function of the box radius 
for non-resonant continuum 
states \cite{DFT84,HT70,MRT93}. 

A more consistent 
treatment for the resonance state is also possible for a spherical 
system. The resonance states are associated with poles of 
the $S$-matrix in the complex energy plane. 
One way to obtain the resonance states is to impose 
the outgoing wave boundary condition in the asymptotic region with 
a complex energy. The state which satisfies this boundary 
condition is referred to as the Gamow state in the literature. 
A similar approach is used in the
complex scaling method, which has been developped originally in atomic
and molecular physics \cite{M98} and subsequently been applied to 
nuclear systems \cite{KLG88,AKI01}. 
Alternatively, the resonance states 
can also be found by monitoring the energy derivative of the phase shift along 
the real energy axis \cite{BLS01}, where 
a maximum appears at the resonance energy. 
For a scattering wave function in 
a spherical potential, a simple factorization 
formula has been derived for its energy and radial dependences 
in the vicinity of an isolated resonance state 
\cite{BLS01,U67,GM70}. 
This factorization property
has been used recently in order to estimate the effect of resonant 
states on pairing correlations in loosely bound nuclei
\cite{SGL00}. 

A resonance structure is much more complicated for a deformed system,  
where several angular momentum components of the single-particle wave function 
are coupled to each other. A standard method to solve this problem 
is to expand the wave function on a discrete basis, which may be deformed 
harmonic oscillator wave functions \cite{V73,GRT90} 
or eigenfunctions of a spherical 
potential\cite{BV80,ZMR03}. 
The positive energy states obtained by diagonalizing the
Hamiltonian in such basis can describe properly
the main properties of the resonant states\cite{HT70,MRT93}, 
but this may require a large number of the basis set. 
A more direct method to obtain resonance states 
in a deformed system is to setup the coupled-channels equations for 
the radial motion and solve them for a complex energy by imposing 
the outgoing wave boundary condition at infinity for all the open 
channels\cite{FML97}. Recently, this method has been successfully applied to 
the proton emission decay of proton-rich nuclei \cite{KBNV00,MFL98}. 
This method has an advantage over the matrix diagonalization method that 
the wave functions are subject only to the radial mesh 
discretization error for the 
Schr\"odinger equation and that the desired asymptotic boundary condition 
can be imposed independently of the size of the mesh interval. 
However, the description of resonant states with such complex wave
functions is generally
difficult to use in nuclear models, since all the evaluated
observables become complex quantities if the rest of the complex
non-resonant states is not taken into account.
We thus prefer to keep the energy as a real variable here. 

The aim of this paper is to develop a practical method to obtain 
scattering wave functions in the vicinity of a multi-channel 
resonance state. A difficulty is 
that there are $N$ linearly independent solutions of the coupled-channels 
equations at a given energy $E$ ($N$ is the number of included
channels), whose asymptotic behavior are all different. For example, a 
physical scattering boundary condition is defined by requiring 
that asymptotically there is an incoming wave 
only in the incident channel. In this case, the equations 
have $N$ degenerate solutions depending on which channel is the incident one. 
Since any linear combination of these $N$ solutions is also a solution 
of the coupled-channels equations, a problem arises as to which combination 
one should take in order to study efficiently 
a multi-channel resonance state. 
If the resonance width is extremely small, as in the case 
of proton decay, one can easily construct a resonance wave function 
by matching to a standing wave solution\cite{ED00,DE01,H01}. 
However, this procedure 
is not applicable to a broad resonance. 

Our method presented in this paper is to start with 
discretized states whose wave functions vanish at a given radius. 
This is an extension of the 
box boundary condition for a spherical system into the deformed regime. 
As in the spherical case, resonances can be identified with those 
states whose energy is stabilized against the box size. 
This method provides the most convenient basis states, where 
the unperturbed states are already a good approximation to the 
true resonance states. In fact, as we will show below, 
the resonance energy thus 
estimated is close to the energy at which the first derivative of the 
eigenphase sum with respect to energy has a maximum. Here, the eigenphase sum 
is defined as the sum of the phase shifts for the eigenchannels, for which 
the $S$- matrix is diagonal\cite{LWDD81}. In the field of 
electron-molecule scattering, it has been known that the eigenphase sum 
has the same energy dependence as 
the elastic phase shift in a one-channel case \cite{M70,H79}. 
Using these unperturbed wave functions, we then employ the first order 
perturbation theory to construct wave functions around a resonance state. 

The paper is organized as follows. In Sec. II, we discuss how to 
generalize the box boundary condition to a multi-channel case. 
This uses a generalized concept of node for a multi-channel 
wave function \cite{J78}. We also discuss the method based on 
the eigenphase sum 
and compare it with the multi-channel box discretization method. 
In Sec. III, we introduce the eigenchannels and discuss their usefulness. 
We consider several cases: the uncoupled (spherical) case, a very narrow 
resonance, and a case where the coupled-channels equations are decoupled. 
In Sec. IV, we discuss the factorization of the wave functions in the 
vicinity of an isolated resonance. We first consider a perturbation to 
the box discretized wave functions for a spherical system in order to 
re-formulate the factorization 
formula of Unger\cite{U67}, which was originally 
derived using 
the $R$-matrix theory \cite{MW69,LT58,V62,BRT83}. 
We then apply it to the eigenbasis 
wave functions, and derive a 
generalization of the factorization formula for a multi-channel wave 
function. We examine the validity of the factorization formula by comparing 
it with the exact solutions of the coupled-channels equations. 
The summary of the paper is then given in Sec. V. In the Appendix, we present 
an explicit form of the coupled-channels equations for a Hamiltonian 
which contains a central and a deformed spin-orbit potentials. 

\section{Single-particle resonances and box-discretized states} 

Let us start with the 
following single-particle (s.p.) Hamiltonian: 
\begin{equation}
H=-\frac{\hbar^2}{2m}\nabla^2+\hat{V}(\vec{r}), 
\end{equation}
where 
the potential $\hat{V}$ may also contain a derivative 
operator such as a spin-orbit interaction. When the potential $\hat V$ 
is non-spherical, the s.p. angular momentum is not conserved and the 
total s.p. wave function has the form 
\begin{equation}
\Psi(\vec{r})=\sum_{ljm}\frac{u_{ljm}(r)}{r}{\cal Y}_{ljm}(\hat{\vec{r}}), 
\label{wftot}
\end{equation}
with 
\begin{equation}
{\cal Y}_{ljm}(\hat{\vec{r}}) = \sum_{m_l,m_s}
\langle l \,m_l \,\frac{1}{2} \,m_s | j \,m\rangle\,
Y_{lm_l}(\hat{\vec{r}})\,\chi_{m_s}. 
\end{equation} 
Here, $Y_{lm_l}$ are spherical harmonics and 
$\chi_{m_s}$ denotes the spin wave function. 
Projecting the Schr\"odinger equation, $H\Psi = E \Psi$, on the 
spin-angular states ${\cal Y}_{ljm}$, the coupled-channels equations 
for the radial wave functions then read,
\begin{equation}
\left[-\frac{\hbar^2}{2m}\frac{d^2}{dr^2}+\frac{l(l+1)\hbar^2}{2mr^2}-E\right]
u_{ljm}(r)+
r\sum_{l'j'm'}\langle{\cal Y}_{ljm}|\hat{V}|
{\cal Y}_{l'j'm'}\rangle\frac{u_{l'j'm'}}{r} = 0.
\label{cc}
\end{equation}

These equations are solved with  certain boundary conditions which 
depend on the problem of interest. 
A standard way to solve them is to generate $N$ linearly independent 
solutions, $N$ being the number of channel states to be included, and then 
to take a linear combination of these $N$ solutions so that the relevant
boundary  condition is satisfied \cite{ED00,HRK99}. 
The linearly independent solutions can be obtained by taking $N$ different
sets of initial  
conditions at $r=0$. We denote these solutions by $\phi_{LL'}(r)$, where $L$ 
refers to the channels while $L'$ refers to a particular choice of initial 
conditions. Here, we use a shorthand notation, $L=(ljm)$. A simple choice for 
the $N$ initial conditions is to impose 
\begin{equation}
\phi_{LL'}(r)\to r^{l+1}\,\delta_{L,L'}~~~~~~~{\rm for}~~~~r\to 0.
\label{initial}
\end{equation}
The coupled-channels equations are solved outwards for each initial 
condition $L'$. The wave functions $u_L(r)$ in Eq.(\ref{cc}) are written in
terms of the  $\phi_{LL'}$ as 
\begin{equation}
u_L(r)=\sum_{L'}C_{L'}\phi_{LL'}(r),
\label{ccwf}
\end{equation}
where the coefficients $C_{L'}$ are determined from the 
asymptotic behavior of $u_L(r)$. 

When the s.p. energy $E$ is positive, the physical wave function is a
scattering wave with an incoming wave behaviour in some particular 
incident channel
$L_0$ and outgoing waves in all channels $L$. Thus,  
the asymptotic boundary condition of the wave function $u_L(r)$ is given 
by
\begin{equation}
u_L(r)\to 
\sqrt{\frac{k}{\pi E}}\,\frac{i}{2}\,
\left\{e^{-i(kr-l\pi/2)}
\delta_{L,L_0}-S_{LL_0}e^{i(kr-l\pi/2)}
\right\}~~~~~~~~~{\rm for}~~~r\to\infty, 
\label{Smat}
\end{equation}
where $k=\sqrt{2mE/\hbar^2}$ and $S_{LL_0}$ is the scattering $S$-matrix. 
The normalization is chosen so that the total wave function 
satisfies 
$\langle \Psi_{E,L_0}|\Psi_{E,L_0'}\rangle = 
\delta(E-E')\delta_{L_0,L_0'}$. 
Here, $|\Psi_{E,L_0}\rangle$ is given by Eq. (\ref{wftot}), 
whose channel components $u_L(r)$ satisfy the boundary condition (\ref{Smat}). 
In practice, the $S$-matrix can be obtained by decomposing $\phi_{LL'}(r)$ 
in the asymptotic region with spherical Hankel functions $h^{(\pm)}_l(kr)$  
and comparing Eq. (\ref{ccwf}) with Eq. (\ref{Smat})\cite{ED00,HRK99}. 

The generalization of the concept of s.p. resonances from the spherical
to the multi-channel cases is most conveniently done by monitoring
the energy dependence of the eigenphase sum \cite{M70,H79}. 
The eigenphases are related to the eigenvalues of the $S$-matrix through 
\begin{equation}
(U^{\dagger}SU)_{aa'}=e^{2i\delta_a}\delta_{a,a'}.
\label{eigen}
\end{equation}
Hazi\cite{H79} has shown that, the sum of the eigenphases  
$\Delta(E)\equiv\sum_a\delta_a(E)$, has the same energy dependence around 
a resonance as the phase shift in a spherical system , i.e., 
\begin{equation}
\Delta(E)= \Delta_0(E) + \tan^{-1}\frac{\Gamma}{2(E_R-E)},
\end{equation}
where the slowly-varying quantity $\Delta_0(E)$ is the sum of the 
background eigenphases, and 
$E_R$ and $\Gamma$ are the multi-channel resonance energy and 
total width, respectively. 
Therefore, the properties of multi-channel resonances can be extracted 
directly 
by plotting $\Delta(E)$, or its energy derivative, as a function of $E$. 
The quantity $\Delta(E)$ is called the eigenphase sum, and it has been 
widely used in the context of electron-molecule scattering (see, e.g., Ref. 
\cite{RIO02} for a recent publication). 

We illustrate the above discussion by the following example. We use a 
Woods-Saxon parametrization for the potential $\hat{V}$ which is given by 
\begin{equation}
\hat{V}(\vec{r})=V_{\rm cent}(\vec{r})+
\nabla(V_{\rm ls}(\vec{r}))
\cdot(-i\nabla\times\vec{\sigma}), 
\label{pot}
\end{equation}
with
\begin{eqnarray}
V_{\rm cent}(\vec{r})&=&V_0(r)-R\,\beta\,\frac{dV_0(r)}{dr}
\,Y_{20}(\hat{\vec{r}}), \\
V_{\rm ls}(\vec{r})&=&V_{\rm so}(r)-R_{\rm so}\,\beta\,
\frac{dV_{\rm so}(r)}{dr}\,Y_{20}(\hat{\vec{r}}), 
\end{eqnarray}
where $V_0(r)$ and $V_{\rm so}(r)$ have a Woods-Saxon shape:
%
\begin{eqnarray}
V_0(r)&=&-V_0/[1+\exp((r-R)/a)], \\
V_{\rm so}(r)&=&V_{\rm so}/[1+\exp((r-R_{\rm so})/a_{\rm so})]. 
\end{eqnarray}
For simplicity, we have expanded the deformed Woods-Saxon potential and kept 
only 
the linear order of the deformation parameter $\beta$\cite{HTD97}. 
Also, we assume 
an axial symmetry, where both the parity $\pi$ and 
the spin projection $K$ onto the $z$ axis 
are conserved. An explicit form of the coupled-channels equations is 
given in the Appendix. 
The parameters of the Woods-Saxon potential are taken to be 
$V_0$=42.0 MeV, $R$ = $R_{\rm so}$ = 1.27 $\times 44^{1/3}$ fm, 
$a=a_{\rm so}$ = 0.67 fm, 
and $V_{\rm so}$= 14.9 MeV/fm$^2$, 
to somehow simulate the neutron potential in the $^{44}$S region. 
As an example we calculate the neutron s.p. levels 
with $K^\pi$ = 5/2$^+$ at $\beta$=0.2. 
We include the $d_{5/2}, 
g_{7/2}, g_{9/2}, i_{13/2}$, and $i_{11/2}$ 
states in the coupled-channels equations. 
Notice that the $s_{1/2}$ and $d_{3/2}$ states do not contribute 
to the $K^\pi=5/2^+$ levels. 
We have checked that the results are not significantly altered 
even if we include 
higher angular momentum components. 

The thick solid line in Fig. 1 
shows the eigenphase sum and its energy derivative as a function
of energy.The contributions from three main eigen-channels 
are also shown by thin lines. 
One clearly sees 
a maximum at $E_R$= 3.44 MeV  
in the first derivative of the eigenphase sum.
The total width can be estimated to be $\Gamma$= 0.46 MeV. 

Let us now compare this continuum result with the s.p. spectrum of a
box-discretized calculation. 
From the structure (\ref{ccwf}) of the general solution, 
it is clear that one can impose the boundary 
condition that the total wave function $\Psi(\vec{r})$ vanishes at 
a radius $R_{\rm box}$. This is equivalent to putting the nucleus in an 
impenetrable spherical box of radius $R_{\rm box}$. 
If the condition  
\begin{equation}
{\rm det}(\phi_{LL'}(R_{\rm box}))=0 
\end{equation}
is satisfied, then 
Eq.(\ref{ccwf}) has a non-trivial solution for $C_L$ such that  
$u_L(R_{\rm box})=0$ for all the channels $L$. 
This is a natural extension of the 
well-known box discretization method for a spherical system. 
We notice that Johnson\cite{J78} has advocated to use  
${\rm det}(\phi(r))$ as a generalized concept of 
node for a multi-channel wave function. We have 
recently used this method to find all the bound state solutions of a deformed 
Skyrme-Hartree-Fock mean field. 

Figure 2 shows s.p. energies 
obtained with the box discretization 
method as a function of the box size $R_{\rm box}$. 
One can clearly see that 
there are two classes of s.p. levels. One consists of those whose 
energy changes significantly as the box radius $R_{\rm box}$ increases, 
and the other contains those whose energy is almost constant as a function of 
$R_{\rm box}$. This behavior is similar to that of box-discretized s.p.
energies for a spherical system where 
the stabilized levels correspond to resonances\cite{DFT84}. 
This example shows that, in the deformed case it is also possible to
generate a complete set of box-discretized states and that some of these
states can be a good approximation to multi-channel s.p. resonances.

\section{Eigenchannel representation}

The states obtained with the box discretization method in the previous 
section are the eigenstates of a potential which 
is the same as the original potential 
for $r<R_{\rm box}$, and infinite for $r \ge R_{\rm box}$. 
They thus form a complete set, and any regular function defined in the
domain $r<R_{\rm box}$ can be expanded on this basis. 
In the next section, we discuss the factorization property of the scattering 
wave function around a resonance 
in terms of this basis set. 
To this end, it appears convenient to introduce the eigenchannel wave
functions \cite{LWDD81} defined in terms of the unitary matrix $U$ of Eq.
(\ref{eigen}),   
\begin{equation}
\tilde{\Psi}_{E,a}(\vec{r})\equiv \sum_{L_0}
\Psi_{E,L_0}(\vec{r})U_{L_0a}.
\end{equation}
We refer to the channels $L$ as the physical 
channels, in order to distinguish them from the eigenchannels $a$. 
Substituting Eqs. (\ref{Smat}) and (\ref{eigen}), one can 
find that the eigenchannel wave functions behave asymptotically as 
\begin{equation}
\tilde{\Psi}_{E,a}(\vec{r})\to 
\frac{1}{r}\sum_L
\sqrt{\frac{k}{\pi E}}\,\frac{i}{2}\,
\left\{e^{-i(kr-l\pi/2)}-e^{2i\delta_a}e^{i(kr-l\pi/2)}
\right\}U_{La}
{\cal Y}_{ljm}(\hat{\vec{r}})
~~~~~~~~~{\rm for}~~~r\to\infty. 
\label{eigenbc}
\end{equation}
For each eigenchannel, the asymptotic 
radial wave functions thus behave in the same way 
for all angular momentum components, and the coupling 
is greatly simplified. Loomba {\it et al.} argued that the eigenchannel wave 
functions provide the most direct way to visualize a resonance state 
\cite{LWDD81}. The eigenchannels were 
also used in the context of the $R$ matrix 
theory \cite{BRT83,DG66,BBD73}. We also mention that the eigenchannel approach 
has been widely used in the field of heavy-ion fusion reactions in 
the context of 
fusion barrier distributions \cite{DLW83,NBT86,RSS91,DHRS98,HTB97}. 
In this section, we discuss the usefulness of this approach for 
multi-channel resonances by considering several cases. 

\subsection{Spherical systems}

In spherical systems, the s.p. angular
momentum is conserved and 
the $S$- matrix is diagonal in the physical channel representation $L$, i.e.,
the physical channels are 
equivalent to the eigenchannels. Likewise, the wave function
$\phi_{LL'}(r)$ with the initial condition (\ref{initial}) 
is diagonal at all the values of $r$, and $C_L$ in Eq. (\ref{ccwf})
is proportional to $\delta_{L,L_0}$ for the incident channel $L_0$. 
From this consideration, it is apparent that the eigenchannel 
wave functions $\tilde{\Psi}_a$ exhibits a resonance behavior for a spherical 
system if the resonance condition is met for an incident channel 
$a=L_0$.  

\subsection{Two-channel decoupled system}

We next consider a special case where the coupled-channels equations 
are completely decoupled by a transformation of the basis. 
For this purpose, we consider the following two-channel 
model \cite{DLW83}, 
\begin{equation}
\left(-\frac{\hbar^2}{2m}\frac{d^2}{dr^2}+V(r)-E\right)
\left(\matrix{
u_1(r) \cr u_2(r)\cr} \right)
+
\left(\matrix{
0 & F(r) \cr 
F(r) & 0 \cr} \right)
\left(\matrix{
u_1(r) \cr u_2(r)\cr} \right)
=0.
\label{2ch}
\end{equation}
This corresponds to a system where the incident channel couples to a 
spinless vibrational state whose 
excitation energy is zero\cite{DLW83,NBT86,HTB97}. This model has been 
used by Dasso {\it et al.} in order to understand the coupling-assisted 
tunneling phenomena in heavy-ion fusion reactions \cite{DLW83}. 

We refer to the basis in the coupled-channels equations (\ref{2ch}) as 
$|1\rangle$ and $|2\rangle$. 
For symmetry reasons the $S$-matrix in this system satisfies 
$S_{11}=S_{22}$ and $S_{12}=S_{21}$, 
and the total wave functions are given by
\begin{eqnarray}
\Psi_1&=&\frac{1}{r}[u_1(r)|1\rangle + u_2(r)|2\rangle], \\
\Psi_2&=&\frac{1}{r}[u_2(r)|1\rangle + u_1(r)|2\rangle]. 
\end{eqnarray}
Here, the asymptotic behaviors of the wave functions $u_1(r)$ and 
$u_2(r)$ are 
\begin{equation}
u_1(r) \to e^{-ikr} - S_{11}\,e^{ikr},
\end{equation}
and 
\begin{equation}
u_2(r) \to  - S_{12}\,e^{ikr},
\end{equation}
respectively. By diagonalizing the $S$ matrix, the eigenchannel wave 
functions read
\begin{eqnarray}
\tilde{\Psi}_1&=&\frac{1}{\sqrt{2}}(\Psi_1+\Psi_2)=\frac{1}{r}
(u_1(r)+u_2(r))\cdot 
\frac{|1\rangle + |2\rangle}{\sqrt{2}}, \\
\tilde{\Psi}_2&=&\frac{1}{\sqrt{2}}(\Psi_1-\Psi_2)=\frac{1}{r}
(u_1(r)-u_2(r))\cdot 
\frac{|1\rangle - |2\rangle}{\sqrt{2}}. 
\end{eqnarray}
Notice that the coupling matrix in the coupled-channels equations (\ref{2ch}) 
can be diagonalized independently of $r$ with the same
unitary operator $U$. Transforming 
Eqs. (\ref{2ch}) with this unitary operator leads to 
\begin{equation}
\left(-\frac{\hbar^2}{2m}\frac{d^2}{dr^2}+V(r)-E\right)
\left(\matrix{
u_1(r)+u_2(r) \cr u_1(r)-u_2(r)\cr} \right)
+
\left(\matrix{
F(r) & 0 \cr 
0 & -F(r) \cr} \right)
\left(\matrix{
u_1(r)+u_2(r) \cr u_1(r)-u_2(r)\cr} \right)
=0.
\end{equation}
The coupled-channels equations are thus decoupled with this basis. 
The radial wave functions for the eigenchannel wave
functions $\tilde{\Psi}_{1,2}$ obeys a one-dimensional Schr\"odinger 
equation with a potential given by $V(r)\pm F(r)$, respectively. 
Therefore, if the one-dimensional potential $V(r)+F(r)$ 
holds a resonance state at an energy $E_R$, the eigenchannel wave
function $\tilde{\Psi}_1$ exhibits a resonance behavior while the 
wave function $\tilde{\Psi}_2$ does not, and vice versa.  
The energy dependence of the wave functions with the physical basis
$\Psi_{1,2}$ is much more complicated. 
Although, in general cases, the coupled-channels equations are not decoupled
if different angular momentum components are coupled to each other, 
it is evident from this example that 
the eigenchannel basis provides a much simpler image of a
multi-channel resonance than the physical basis. 

\subsection{Very narrow resonance}

As a final example, we consider a case where the resonance width is 
very small, as in proton or alpha decays. In
Refs. \cite{ED00,DE01,H01}, the energy of a multi-channel resonance 
for a proton emission decay was found by imposing the asymptotic 
boundary condition that the 
channel wave functions are proportional to $G_l(kr)$ 
for all the channels, where $G_l(kr)$ is the irregular Coulomb 
wave function. From Eq. (\ref{eigenbc}), one can see that this state 
should correspond to one of the eigenbasis wave functions whose
eigenphase is $\pi/2$. In this subsection, we would like to 
demonstrate that 
the total width is exhausted by only one eigenchannel 
when it is very small. 

To this end, let us start with the Breit-Wigner expression 
for a multi-channel $S$-matrix \cite{MW69,LT58,V62,BRT83}. 
If there is only one isolated resonance of energy $E_R$ 
in the energy region of interest, 
the $S$-matrix can be aproximated 
in the vicinity of this resonance as 
\begin{equation}
S_{LL'}=e^{2i\phi_L}\delta_{L,L'}-e^{i(\phi_L+\phi_{L'})}
\frac{i\sqrt{\Gamma_L\Gamma_{L'}}}{E-E_R+i\Gamma/2},
\end{equation}
where $\phi_L$ is the background phase shift and $\Gamma$ is the total
width given as a sum of partial widths $\Gamma_L$. For a very narrow 
resonance, the background phase shifts can often be neglected. 
The left hand side of Eq. (\ref{eigen}) then reads 
\begin{equation}
(U^{\dagger}SU)_{aa'}=\delta_{a,a'}-
\frac{i\sqrt{\Gamma_a\Gamma_{a'}}}{E-E_R+i\Gamma/2},
\label{eigen2}
\end{equation}
where $\sqrt {\Gamma_a}=\sum_L\sqrt {\Gamma_L}U_{La}$ is 
related to 
the partial width for the 
eigenchannel $a$. By definition of the unitary operator $U$, the
off-diagonal components of the right hand side of Eq. (\ref{eigen2}) 
should vanish. The only possibility for this to be satisfied is 
that $\Gamma_a$ is zero except for one particular channel $a_0$. 
Thus, around the resonance the eigenphase shifts are given by 
$\delta_{a_0}=\tan^{-1}[\Gamma/(2(E_R-E))]$ for the eigenchannel
$a_0$, and $\delta_a=0$ for the other channels. Again, this example 
clearly shows that the eigenchannel representation provides the most direct 
manifestation of a multi-channel resonance state. 

\section{Factorization of scattering wave functions}

We now consider the most general case for a multi-channel resonance,
i.e., when the resonance width is not very small.
For a spherical system having an isolated 
single-particle resonance at energy $E_R$, 
Unger \cite{U67} has shown that the scattering
radial wave function at energy $E$  whose asymptotic form is 
\begin{equation}
\psi_E(r)\to \sqrt{\frac{k}{\pi E}}e^{i\delta(E)}\sin(kr-l\pi/2+\delta(E))
~~~~~~~{\rm for}~~~~r\to \infty,
\label{asymptotic}
\end{equation}
can be approximately factorized into a product of an $r$-independent
function $F(E)$ times the wave function at resonance energy. This
factorization property holds if 
\begin{equation} 
\vert E-E_R \vert \ll \vert V({\bf r})\vert,
\label{condition}
\end{equation}
a condition which is realized generally in the internal region of
the potential when $\vert E-E_R \vert$ is less than the resonance width
$\Gamma$. In Ref.\cite{U67} the function $F(E)$ is written in terms of a
Breit-Wigner function, but for an isolated resonance it is easy to relate
this to the derivative of the phase shift with respect
to energy, so that the factorization property can be expressed as
\begin{equation}
\psi_E(r)\sim 
e^{i\delta(E)}\sqrt{\frac{1}{\pi}\frac{d\delta}{dE}}\,
\frac{\psi_{E_R}(r)}{\left[\int^R_0dr\,|\psi_{E_R}(r)|^2\right]^{1/2}}, 
\label{Unger}
\end{equation}
under the condition (\ref{condition}). Thus, the factorization function
$F(E)$ is just $e^{i\delta(E)}\sqrt{\frac{1}{\pi}\frac{d\delta}{dE}}$ except
for a normalization factor. 
In this section, we generalize this formula to a deformed system.
We do not attempt to give an analytic derivation, which 
could be done by using multi-channel Jost functions \cite{Newton}, 
but we would like to have a
hint on how should be 
the generalized form of Eq.(\ref{Unger}). To this end
we find it useful to use
the box discretized states of Sec. II to express the function $F(E)$, first
in the familiar spherical case, then in the general case.

\subsection{Spherical system}

The infinite set of all discretized states $X_\lambda(r)$ with corresponding
eigen-energies $E_\lambda$ calculated in a box of radius $R_{\rm box}$ form a
complete orthonormal set, and one can expand on this basis any regular
function defined in the domain $[0, R_{\rm box}[$ (the domain excludes
$r=R_{\rm box}$). In particular, the inner part of the scattering wave 
function $\psi_E(r)$ can be expanded as 
\begin{equation}
\psi_E(r)=\sum_\lambda C_\lambda(E) X_\lambda(r)~~~r \in [0, R_{\rm box}[~,
\label{expansion}
\end{equation}
with 
\begin{equation}
C_\lambda(E) = \int_0^{R_{\rm box}}dr\,X_\lambda(r)\psi_E(r).
\end{equation}
Notice that this expansion does not hold at $r=R_{\rm box}$ in general. 
Here and in the following, all integrals $\int_0^{R_{\rm box}}dr$ 
are meant to be $\lim_{\epsilon \to 0}\int_0^{R_{\rm box}-
\epsilon}dr$. 
With this definition, 
the expansion (\ref{expansion}) is still meaningful even when 
the radius $r$ is infinitesimally smaller than $R_{\rm box}$. 
Since the wave functions $X_\lambda(r)$ and $\psi_E(r)$ obey the
same Schr\"odinger equation (but with different energies from each other), 
\begin{eqnarray}
&&-\frac{\hbar^2}{2m}\frac{d^2\psi_E}{dr^2}
+\left(V(r)+\frac{l(l+1)\hbar^2}{2mr^2}-E\right)\psi_E=0~, \nonumber \\
&&-\frac{\hbar^2}{2m}\frac{d^2X_\lambda}{dr^2}+
\left(V(r)+\frac{l(l+1)\hbar^2}{2mr^2}-E_\lambda\right)X_\lambda=0~. 
\end{eqnarray}
the expansion coefficients $C_\lambda(E)$ are given by 
\begin{eqnarray}
C_\lambda(E)&=&
\frac{1}{E-E_\lambda} \frac{\hbar^2}{2m}
\int^{R_{\rm box}}_0dr\,\left(\psi_E\frac{d^2X_\lambda}{dr^2}
-X_\lambda\frac{d^2\psi_E}{dr^2}\right)~, \nonumber \\
&=&
\frac{1}{E-E_\lambda} \frac{\hbar^2}{2m}
\psi_E(R_{\rm box})X'_\lambda(R_{\rm box})~. 
\end{eqnarray}
In the last step we have used the boundary condition 
$X_\lambda(R_{\rm box})=0$.  

In the vicinity of $E_R \simeq E_{\lambda_R}$, the contribution from 
$\lambda=\lambda_R$ dominates 
the expansion (\ref{expansion}).  Retaining only this term 
in the expansion 
(this corresponds to the one term approximation in the $R$-matrix
theory \cite{MW69,LT58,V62,BRT83}), the scattering wave function
around the resonance and inside the potential barrier reads 
\begin{equation}
\psi_E(r)\sim 
\frac{1}{E-E_{\lambda_R}}\cdot \frac{\hbar^2}{2m}
\psi_E(R_{\rm box})X'_{\lambda_R}(R_{\rm box})X_{\lambda_R}(r).
\label{box0}
\end{equation}
We now set the $R$-matrix radius $R$ to be 
$R_{\rm box}$, approximate $X_{\lambda_R}(r)$ by the resonance wave function
$\psi_{E_R}(r)$ (except for the normalization) and finally  
assume that $kR_{\rm box}$ is sufficiently large 
so that the 
$\psi_E(r)$ at $R_{\rm box}$ can be replaced by
its asymptotic form (\ref{asymptotic}). Then,  
a comparison between Eqs. (\ref{Unger}) and (\ref{box0}) leads to 
\begin{equation}
e^{i\delta(E)}\sqrt{\frac{1}{\pi}\frac{d\delta}{dE}}
=
\frac{1}{E-E_{\lambda_R}}\cdot \frac{\hbar^2}{2m}
\sqrt{\frac{k}{\pi E}}\,e^{i\delta(E)}\sin(kR_{\rm box}-l\pi/2+\delta(E))
\frac{X'_{\lambda_R}(R_{\rm box})}
{\left[\int^{R_{\rm box}}_0dr\,\vert X_{\lambda_R}(r)\vert^2\right]^{1/2}}~. 
\label{factorization0}
\end{equation}
We have thus established the relation between the function $F(E)$ (except
for a normalization factor) and the derivative of the discretized state
$X_{\lambda_R}$ at $r=R_{box}$. This relation is our guideline for
obtaining the generalized function $F(E)$ in the deformed case.
Notice that we have explicitly written down the normalization of
$X_{\lambda_R}(r)$ so that Eq. (\ref{factorization0}) can be used also
when the function $X_{\lambda_R}(r)$ is not normalized.   

\subsection{Deformed system}

The procedure is very similar to that of the
spherical case. 
Using the complete set of discretized wave 
functions introduced in Sec.II : 
\begin{equation}
X_\lambda(\vec{r})=\sum_L\frac{X_{L\lambda}(r)}{r} 
{\cal Y}_{L}(\hat{\vec{r}}), 
\end{equation}
we expand the eigenchannel wave functions
$\tilde{\Psi}_{E,a}(\vec{r})$:  
\begin{eqnarray}
\tilde{\Psi}_{E,a}(\vec{r})& \equiv &\sum_L\frac{u^{(a)}_L(r)}{r} 
{\cal Y}_{L}(\hat{\vec{r}})~, \nonumber \\
 & = & \sum_\lambda C_\lambda(E) X_\lambda(\vec{r}), 
\label{expansion2}
\end{eqnarray}
with 
\begin{equation}
C_\lambda = \sum_L \int_0^{R_{\rm box}}dr\,X_{L\lambda}(r)u^{(a)}_L(r)~.
\end{equation}
The coupled-channels equations (\ref{cc}) lead to 
\begin{equation}
C_\lambda(E)=
\frac{1}{E-E_\lambda}\cdot \frac{\hbar^2}{2m}
\sum_L u^{(a)}_L(R_{\rm box})
X'_{L\lambda}(R_{\rm box})~. 
\label{eqq}
\end{equation}
As in the spherical system, we retain only the contribution from
$\lambda=\lambda_R$ in the expansion (\ref{expansion2}) for
$E \simeq E_{\lambda_R}$.
In the asymptotic region (for a large $R_{\rm box}$) the wave function
$u^{(a)}_L(r)$ has exactly the same asymptotic form (except for a
factor $U_{La}$) as the scattering wave function $\psi_E(r)$ in the spherical
system (see Eqs. (\ref{eigenbc}) and (\ref{asymptotic})), the only
difference being that the spherical phase shift $\delta(E)$ is replaced by
the eigenphase $\delta_a$.
Thus, we can replace in Eq.(\ref{eqq}) $u^{(a)}_L(R_{\rm box})$ by
$\psi_E(R_{\rm box}) U_{La}$ with $\psi_E$ represented by its asymptotic
form (\ref{asymptotic}). Comparing the resulting expression with 
Eq. (\ref{factorization0}), 
we obtain the desired 
factorization formula for a multi-channel resonance,  
\begin{equation}
C_{\lambda_R}\sim 
e^{i\delta_a(E)}\sqrt{\frac{1}{\pi}\frac{d\delta_a}{dE}}\cdot
\left(
\sum_L 
\left[\int^{R_{\rm box}}_0dr\,\vert X_{L\lambda_R}(r)\vert
^2\right]^{1/2}U_{La}\right)~. 
\label{deffactor2}
\end{equation}
Here, $\sum_L 
\left[\int^{R_{\rm box}}_0dr\,X_{\lambda_R}(r)^2\right]^{1/2}U_{La}$
is the probability of finding the particle inside the box for the
eigenchannel $a$. 

In practice, one may replace the box wave function $X_{\lambda_R}$ in
Eq. (\ref{deffactor2}) by the scattering wave 
functions at the resonance, for each eigenchannel $a$, i.e., 
\begin{equation}
\tilde{\Psi}_{E,a}(\vec{r})\sim C_{a}(E)\tilde{\Psi}_{E_R,a}(\vec{r})
/C_{a}(E_R)
\label{deffactor3}
\end{equation}
with
\begin{equation}
C_a(E)
=
\sqrt{\frac{1}{\pi}\frac{d\delta_a}{dE}}\cdot
\frac
{\sum_L 
\left(\left[\int^{R_{\rm box}}_0dr\,|u^{(a)}_L(r)|^2\right]^{1/2}
U_{La}(E)\right)}
{ 
\left[\sum_L \int^{R_{\rm box}}_0dr\,|u^{(a)}_L(r)|^2\right]^{1/2}}, 
\label{deffactor4}
\end{equation}
where the wave functions $u^{(a)}_L(r)$ are evaluated at the resonance 
energy $E_R$. One may also choose the phase factor of 
the eigenchannel wave functions $\tilde{\Psi}_{E,a}$ for each channel $a$ 
and energy $E$ so that the 
factor $e^{i\delta_a}$ does not appear in Eq. (\ref{deffactor4}). 
With this, the wave functions become real numbers. 

Figure 3 shows the applicability of the factorization formula for the system 
considered in Figs. 1 and 2. At the resonance energy, $E=E_R=3.44$ MeV, 
the eigenphase shifts $\delta_a/\pi$ are calculated as 0.559,
2.53$\times 10^{-3}$, 4.12$\times 10^{-2}$, 1.06$\times 10^{-5}$, and 
2.79$\times 10^{-5}$ for the five eigenchannels (see the thin lines 
in Fig. 1). 
Therefore, there is one eigenchannel which
dominates the eigenphase sum $\Delta(E)$ around the resonance. 
In the figure, 
the $d_{5/2}$, $g_{7/2}$, and $g_{9/2}$ wave function components 
are shown for 
this particular eigenchannel. The exact wave functions at the
resonance energy are denoted by the thick solid line, while those at
$E=3.54$ MeV are given by the thin solid line. The dashed line is the 
result of the factorization formula (\ref{deffactor3}) and 
(\ref{deffactor4}). We take $R_{\rm box}=20$ fm to evaluate the
integrations in the formula. 
For the $l=4$ components, one clearly sees that the energy dependence,
as well as the 
radial dependence for $r\le 10$ fm, of the radial wave functions are 
well expressed by the factorization formula. 
We have confirmed 
that the agreement is 
even better for the $l=6$ components (not shown). 
For the $l=2$
component, the wave function is not localized inside the barrier, 
and the factorization formula does not work well. 
As expected, the factorization approximation is better for
high $l$ states, which are more localized by the centrifugal
barrier. 

Figure 4 shows the spin projected total wave function (multiplied by $r$), 
\begin{equation}
\langle \chi_{1/2}|r\,\tilde{\Psi}_{E,a}(\vec{r})\rangle
= \sum_{lj}u^{(a)}_{lj}(r)\cdot 
\langle l,\,K-\frac{1}{2},\,\frac{1}{2},\,\frac{1}{2}|j\,K\rangle 
Y_{l\,K-1/2}(\hat{\vec{r}}),
\end{equation}
for this eigenchannel $a$. The upper and the lower panes are the wave
functions as a function of $r$ for $(\theta,\phi)=(\pi/4,0)$ and 
$(\theta,\phi)=(\pi/2,0)$, respectively. (Note that 
the total wave function vanishes along $\theta=0$ for $K=5/2$.) 
As we have shown in fig. 3, the factorization formula works reasonably well, 
especially inside the potential barrier, i.e., $r\leq$ 4.23 fm. 

In Fig. 5, we show more in detail the energy dependence of the main component 
of the multi-channel wave function (i.e., the $g_{9/2}$ component) 
around the resonance energy, 
and its comparison with the factorization method. 
At $E=E_R+0.1\Gamma$, 
the factorization formula works, and the exact radial dependence 
of the wave function 
is well reproduced, as we already showed in Fig. 3. 
As the energy increases, the performance of the factorization becomes worse. 
At $E=E_R+\Gamma$, the factorization still works 
inside the potential, but the deviation becomes significant in the 
outside (see the 
middle panel). By the energy $E_R+2\Gamma$, the wave function does not 
resemble the resonance wave function, and the factorization formula looses 
its applicability, as was already known for the spherical system\cite{GM70}. 

In Ref. \cite{LM70}, Lejeune and Mahaux used a solvable two-level
model to argue that the energy dependence of the 
wave function components is quite different from each other and 
a simple scaling does not hold. However, this does not come out in 
the present study. The energy dependences are similar for all the 
angular momentum components, except for the $d$-wave which shows a
non-resonant behavior. This different conclusion may come from the 
fact that here the calculations are done with 
the eigenchannel basis, in contrast to Lejeune and Mahaux, who worked 
with the physical basis. 
This fact again suggests that the eigenchannel 
basis provides a powerful means to study a multi-channel resonance. 

\section{Summary}

We have used the eigenchannel representation of the multichannel wave function 
to investigate the properties of a single-particle resonance in a deformed 
system. We first showed that the box boundary condition can be generalized 
to deformed systems if one uses the generalized concept of the 
wave function node,  
that is to impose the condition that the 
determinant of the wave function matrix vanishes at the boundary of a 
spherical box. We have solved the coupled-channels equations with this 
boundary condition and showed that the discretized energy is nearly 
a constant as a function of the box size for a multichannel resonance state. 
The energy $E_R$ 
and the width $\Gamma$ 
of the resonant state were also calculated by using  the
energy dependence of the eigenphase sum. 
We have next studied the properties of the eigenbasis which  
diagonalizes the $S$-matrix. We argued that the eigenchannel representation 
has the most direct connection to a multi-channel resonance state. 
Using this representation, we have 
derived a formula which relates the wave functions 
at different energies around a resonance state, and studied its applicability. 
For an isolated resonance,  
we have observed a nice agreement between the approximate 
and exact wave functions inside the potential radius at energies 
$|E-E_R| \lesssim \Gamma$. 

In many-body problems, one often takes the single-particle states as a good 
building block to start with. This is the case, for instance, in the 
BCS calculation for a pairing Hamiltonian. If one includes unbound 
states in the calculation, one then 
has to deal with integrals over a wide range 
of energies in order to compute matrix elements. In the vicinity of a narrow 
resonance, the energy interval has to be taken small, which may be quite 
time consuming. 
This difficulty is substantially lessened if one uses the factorization 
formula derived in this article. 
Work is now in progress which uses this formula 
for resonance contributions in a Hartree-Fock plus BCS calculation for 
deformed nuclei.  
We will report the results in a separate paper. 

\section*{acknowledgments}

We thank G.F. Bertsch, M. Grasso, E. Khan, J. Libert, 
N. Sandulescu, P. Schuck, M. Urban, and N. Vinh Mau 
for useful discussions,  
and K. Kato for drawing our attention to 
Ref. \cite{J78}. 
K.H. thanks the Theory group of IPN Orsay for its warm hospitality, 
where this work was done, and the Kyoto University Foundation for 
financial support.

\begin{appendix}
\section{Coupled-channels equations with a deformed spin-orbit potential}

In this Appendix, we present an explicit form of the coupling matrix 
elements in the coupled-channels equations, Eq. (\ref{cc}), for a 
Hamiltonian with a deformed spin-orbit interaction in the form of 
Eq.(\ref{pot}). We first perform the multipole decomposition for the 
potentials: 
\begin{eqnarray}
V_{\rm cent}(\vec{r})&=&\sum_{\lambda,\mu}
V^{(\lambda\mu)}_{\rm cent}(r)Y_{\lambda\mu}(\hat{\vec{r}}), \\
V_{\rm ls}(\vec{r})&=&\sum_{\lambda,\mu}
V^{(\lambda\mu)}_{\rm ls}(r)Y_{\lambda\mu}(\hat{\vec{r}}).
\end{eqnarray}

We use the Wigner-Eckart theorem 
\cite{E57} to evaluate the matrix elements. 
For the central potential $V_{\rm cent}$, 
this yields\cite{RG92}
\begin{eqnarray}
&& r\langle{\cal Y}_{ljm}|\hat{V}_{\rm cent}|
{\cal Y}_{j'l'm'}\rangle\frac{u_{j'l'm'}(r)}{r} \nonumber \\
&=&
\sum_{\lambda,\mu}
(-)^{j-m}
\left(\matrix{
j & \lambda & j' \cr 
-m & \mu & m' \cr} \right)
\langle{\cal Y}_{jl}||Y_\lambda||
{\cal Y}_{j'l'}\rangle \,
V^{(\lambda\mu)}_{\rm cent}(r)u_{j'l'm'}(r), \\
&=&\sum_{\lambda,\mu}
(-)^{j-m}
\left(\matrix{
j & \lambda & j' \cr 
-m & \mu & m' \cr} \right)
\,(-)^{\frac{1}{2}+j}\,
\frac{\hat{j}\hat{\lambda}\hat{j'}}{\sqrt{4\pi}}
\left(\matrix{
j & \lambda & j' \cr 
1/2 & 0 & -1/2 \cr} \right)\,
V^{(\lambda\mu)}_{\rm cent}(r)u_{j'l'm'}(r),
\end{eqnarray}
where $\hat{l}$ is defined by $\sqrt{2l+1}$. 

For the spin-orbit part, we first decompose it as\cite{E57}
\begin{eqnarray}
\nabla(V_{\rm ls}(\vec{r}))
\cdot(-i\nabla\times\vec{\sigma})
&=&
\sum_{\lambda,\mu}
\left\{
-\sqrt{\frac{\lambda+1}{2\lambda+1}}
\left(\frac{dV^{(\lambda\mu)}_{\rm ls}}{dr}-\frac{\lambda}{r}
V^{(\lambda\mu)}_{\rm ls}(r)\right)
[Y_{\lambda+1}(-i\nabla\times\vec{\sigma})]^{(\lambda\mu)}\right. \nonumber \\
&&\left.+\sqrt{\frac{\lambda}{2\lambda+1}}
\left(\frac{dV^{(\lambda\mu)}_{\rm ls}}{dr}+\frac{\lambda+1}{r}
V^{(\lambda\mu)}_{\rm ls}(r)\right)
[Y_{\lambda-1}(-i\nabla\times\vec{\sigma})]^{(\lambda\mu)}\right\}, 
\end{eqnarray}
and use the Wigner-Eckart theorem for the 
operators $[Y_{\lambda\pm 1}(-i\nabla\times\vec{\sigma})]^{\lambda\mu}$. 
That is, 
\begin{eqnarray}
&& r\langle{\cal Y}_{ljm}|
\nabla(V_{\rm ls}(\vec{r}))
\cdot(-i\nabla\times\vec{\sigma})|
{\cal Y}_{j'l'm'}\rangle\frac{u_{j'l'm'}(r)}{r} \nonumber \\
&=&
\sum_{\lambda,\mu}
(-)^{j-m}
\left(\matrix{
j & \lambda & j' \cr 
-m & \mu & m' \cr} \right) \nonumber \\
&& \times 
\left\{
-\sqrt{\frac{\lambda+1}{2\lambda+1}}
\left(\frac{dV^{(\lambda\mu)}_{\rm ls}}{dr}-\frac{\lambda}{r}
V^{(\lambda\mu)}_{\rm ls}(r)\right)
r \langle{\cal Y}_{jl}||
[Y_{\lambda+1}(-i\nabla\times\vec{\sigma})]^{(\lambda)}||
{\cal Y}_{jl}\rangle 
\frac{u_{j'l'm'}(r)}{r}\right. \nonumber \\
&&\left.+\sqrt{\frac{\lambda}{2\lambda+1}}
\left(\frac{dV^{(\lambda\mu)}_{\rm ls}}{dr}+\frac{\lambda+1}{r}
V^{(\lambda\mu)}_{\rm ls}(r)\right)
r \langle{\cal Y}_{jl}||
[Y_{\lambda-1}(-i\nabla\times\vec{\sigma})]^{(\lambda)}||
{\cal Y}_{jl}\rangle 
\frac{u_{j'l'm'}(r)}{r}
\right\}. 
\end{eqnarray}
The reduced matrix elements are given by \cite{RG92}
\begin{eqnarray}
\langle {\cal Y}_{jl}||
&& [Y_{\lambda\pm 1}(-i\nabla\times\vec{\sigma})]^{(\lambda)}
||{\cal Y}_{j'l'}\rangle \nonumber \\
&&=
\sum_{j''l''}(-)^{\lambda+j+j'+1}\sqrt{36}\,\hat{\lambda}
\hat{j''}\hat{j'}
\left\{
\matrix{
\lambda\pm 1 & 1 & \lambda \cr 
j' & j & j'' \cr} \right\}\, \nonumber \\
&& \times \left\{\matrix{
l'' & l' & 1 \cr 
1/2  & 1/2 & 1 \cr
j'' & j' & 1} \right\}
\langle{\cal Y}_{jl}||Y_\lambda||
{\cal Y}_{j''l''}\rangle 
\langle l''||\nabla ||l' \rangle,
\end{eqnarray}
with 
\begin{equation}
\langle l''||\nabla ||l' \rangle = \left\{
\matrix{\sqrt{l'+1}\left(\frac{d}{dr}-\frac{l'}{r}\right) 
& {\rm if}~~l''=l'+1, \cr
-\sqrt{l'}\left(\frac{d}{dr}+\frac{l'+1}{r}\right) 
& {\rm if}~~l''=l'-1. \cr}\right.
\end{equation}
This yields 
\begin{eqnarray}
&&r \langle{\cal Y}_{jl}||
[Y_{\lambda\pm 1}(-i\nabla\times\vec{\sigma})]^{(\lambda)}||
{\cal Y}_{jl}\rangle 
\frac{u_{j'l'm'}(r)}{r} \nonumber \\
&&=
\sum_{j''=l'+1/2}^{l'+3/2}
(-)^{\lambda+j+j'+1}\sqrt{36}\,\hat{\lambda}
\hat{j''}\hat{j'}
\left\{
\matrix{
\lambda\pm 1 & 1 & \lambda \cr 
j' & j & j'' \cr} \right\}\, \nonumber \\
&& \times 
\left\{\matrix{
l'+1 & l' & 1 \cr 
1/2  & 1/2 & 1 \cr
j'' & j' & 1} \right\}
\langle{\cal Y}_{jl}||Y_\lambda||
{\cal Y}_{j''l'+1}\rangle 
\cdot \sqrt{l'+1}\left(\frac{d}{dr}-\frac{l'+1}{r}\right)u_{j'l'm'}(r) 
\nonumber \\ 
&&+
\sum_{j''=l'-1/2}^{l'-3/2}
(-)^{\lambda+j+j'+1}\sqrt{36}\,\hat{\lambda}
\hat{j''}\hat{j'}
\left\{
\matrix{
\lambda\pm 1 & 1 & \lambda \cr 
j' & j & j'' \cr} \right\}\, \nonumber \\
&& \times 
\left\{\matrix{
l'-1 & l' & 1 \cr 
1/2  & 1/2 & 1 \cr
j'' & j' & 1} \right\}
\langle{\cal Y}_{jl}||Y_\lambda||
{\cal Y}_{j''l'-1}\rangle 
\cdot (-\sqrt{l'})\left(\frac{d}{dr}+\frac{l'}{r}\right)u_{j'l'm'}(r). 
\end{eqnarray}
The matrix elements of the coupling potential were evaluated in this way 
in Ref. \cite{H01} for the vibrational excitation in the 
proton decay of spherical nuclei. 
 
Notice that the matrix elements become trivial for the monopole 
part ($\lambda=0$), 
and are given by 
\begin{eqnarray}
&&
r\langle{\cal Y}_{ljm}|\hat{V}_{\lambda=0}|
{\cal Y}_{j'l'm'}\rangle\frac{u_{j'l'm'}(r)}{r} \nonumber \\
&&
=
\left[\frac{V^{(00)}_{\rm cent}(r)}{\sqrt{4\pi}}
+\frac{1}{\sqrt{4\pi}r}\frac{dV^{(00)}_{\rm ls}(r)}{dr}
\left(j(j+1)-l(l+1)-\frac{3}{4}\right)\right]\,u_{ljm}(r)\,
\delta_{j,j'}\delta_{l,l'}\delta_{m,m'}.
\end{eqnarray}
\end{appendix}

\begin{figure}
  \begin{center}
    \leavevmode
    \parbox{0.9\textwidth}
           {\psfig{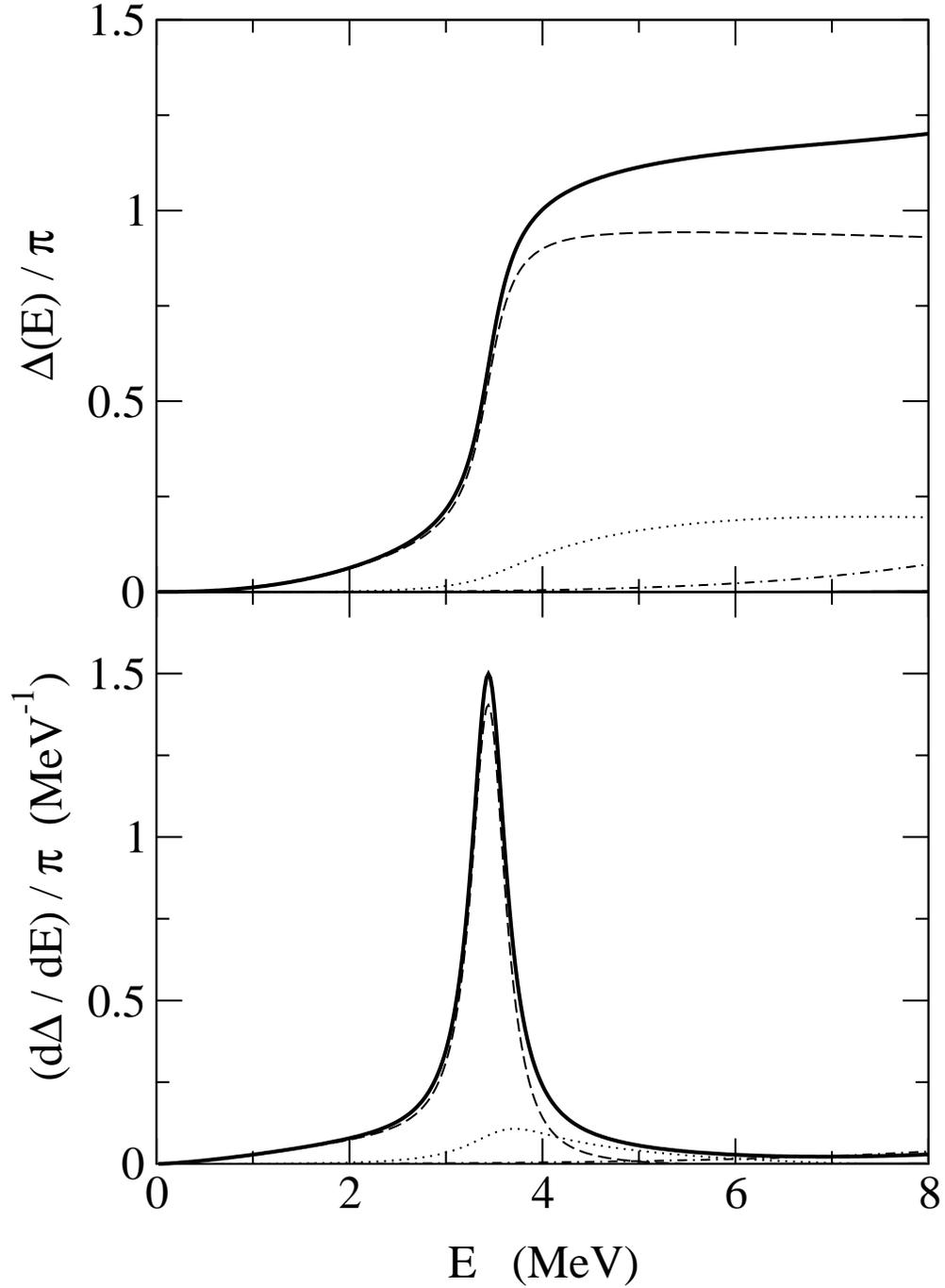}}
  \end{center}
\protect\caption{
The eigenphase sum (the upper panel) and its energy derivative (the lower
    panel) as a function of energy for neutron positive energy states 
of $^{44}$S at $\beta$ = 0.2. 
The spin projection onto the 
symmetry axis and the parity are taken to be $K^\pi=5/2^+$. 
These are denoted by the thick solid line, while the contributions from
    three main eigenchannesl are also shown by the thin dashed, dotted
    and dot-dashed lines. 
}
\end{figure}

\begin{figure}
  \begin{center}
    \leavevmode
    \parbox{0.9\textwidth}
           {\psfig{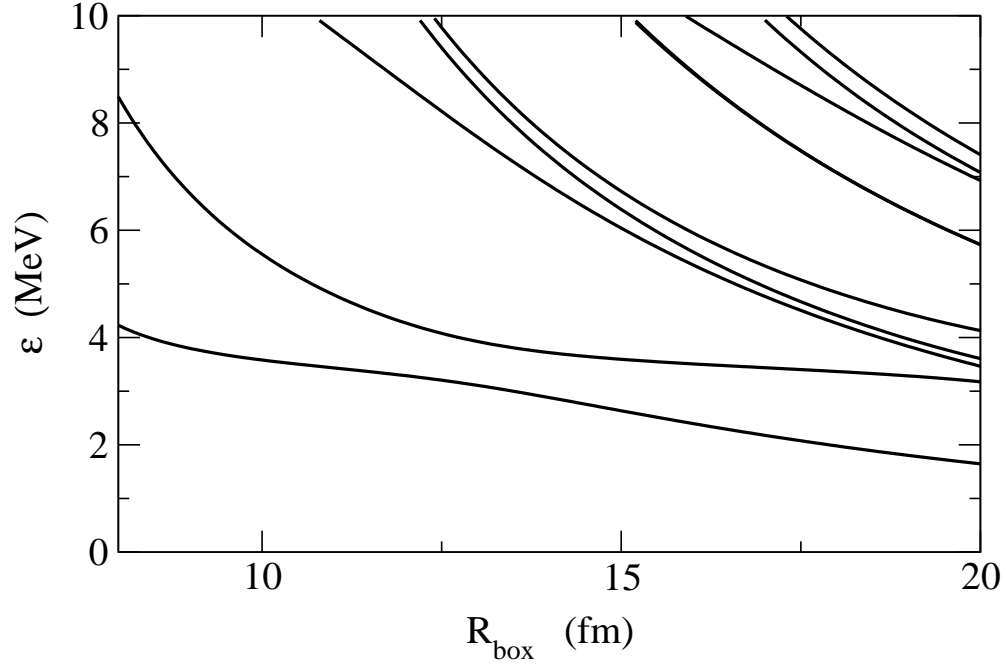}}
  \end{center}
\protect\caption{
Neutron single-particle energies of $^{44}$S 
obtained with the box discretization 
method as a function of the box radius $R_{\rm box}$. 
The axial symmetry is assumed, 
with the deformation parameter $\beta$ of 0.2. The spin projection onto the 
symmetry axis and the parity are taken to be $K^\pi=5/2^+$. 
The potential parameters are the same as in fig. 1. 
}
\end{figure}

\begin{figure}
  \begin{center}
    \leavevmode
    \parbox{0.9\textwidth}
           {\psfig{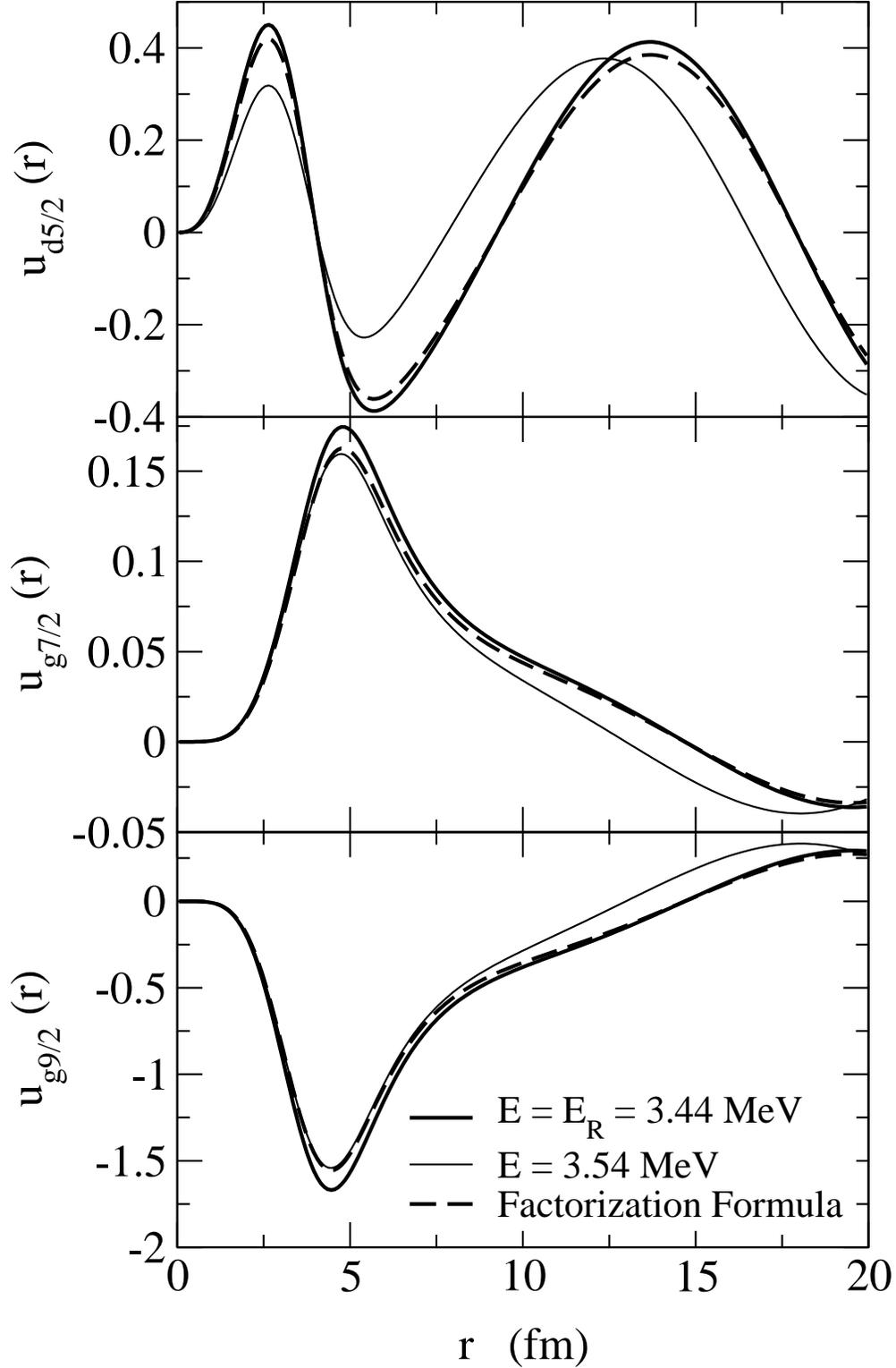}}
  \end{center}
\protect\caption{
The channel wave functions around the neutron resonance state 
of $^{44}$S at $\beta$ = 0.2 and for $K^\pi=5/2^+$. 
Those are for the eigenchannel basis which dominates the total width. 
The thick and thin solid lines are the wave functions at the resonance
    energy, $E_R=3.44$ MeV and at $E$=3.54 MeV, respectively. The
    dashed line is obtained by using the factorization formula for a
    multi-channel resonance state. 
The potential parameters are the same as in fig. 1. 
}
\end{figure}

\begin{figure}
  \begin{center}
    \leavevmode
    \parbox{0.9\textwidth}
           {\psfig{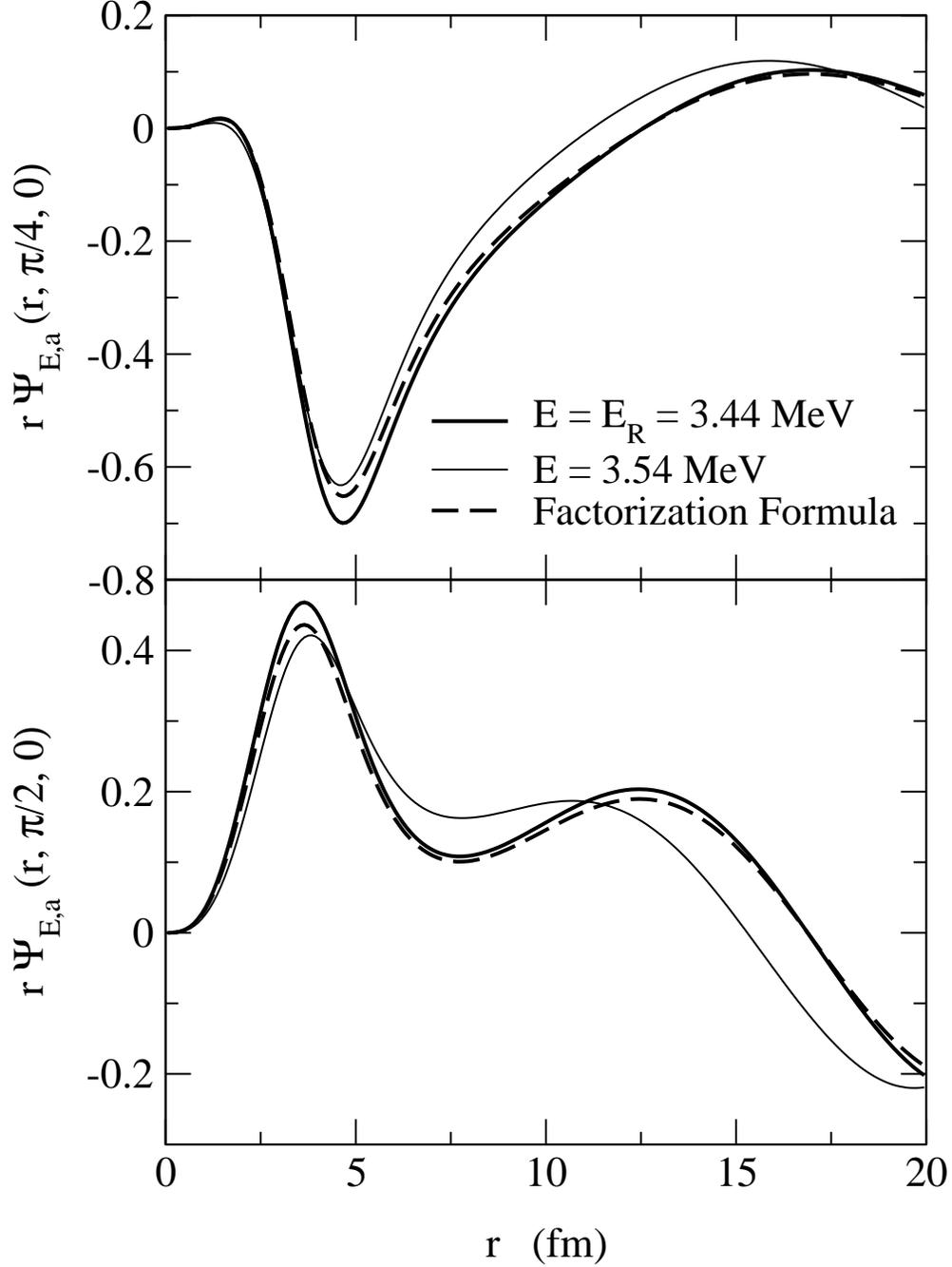}}
  \end{center}
\protect\caption{
The total eigenchannel wave function around the neutron resonance state 
of $^{44}$S at $\beta$ = 0.2 and for $K^\pi=5/2^+$. 
The spin degree of freedom is eliminated by projecting the wave
function onto the upper spin component. 
The upper and the lower panels are for the wave function along the 
$(\theta,\phi)=(\pi/4,0)$ and the $(\theta,\phi)=(\pi/2,0)$
    directions, respectively. The meaning of each line is the same as
    in fig. 3. }
\end{figure}

\begin{figure}
  \begin{center}
    \leavevmode
    \parbox{0.9\textwidth}
           {\psfig{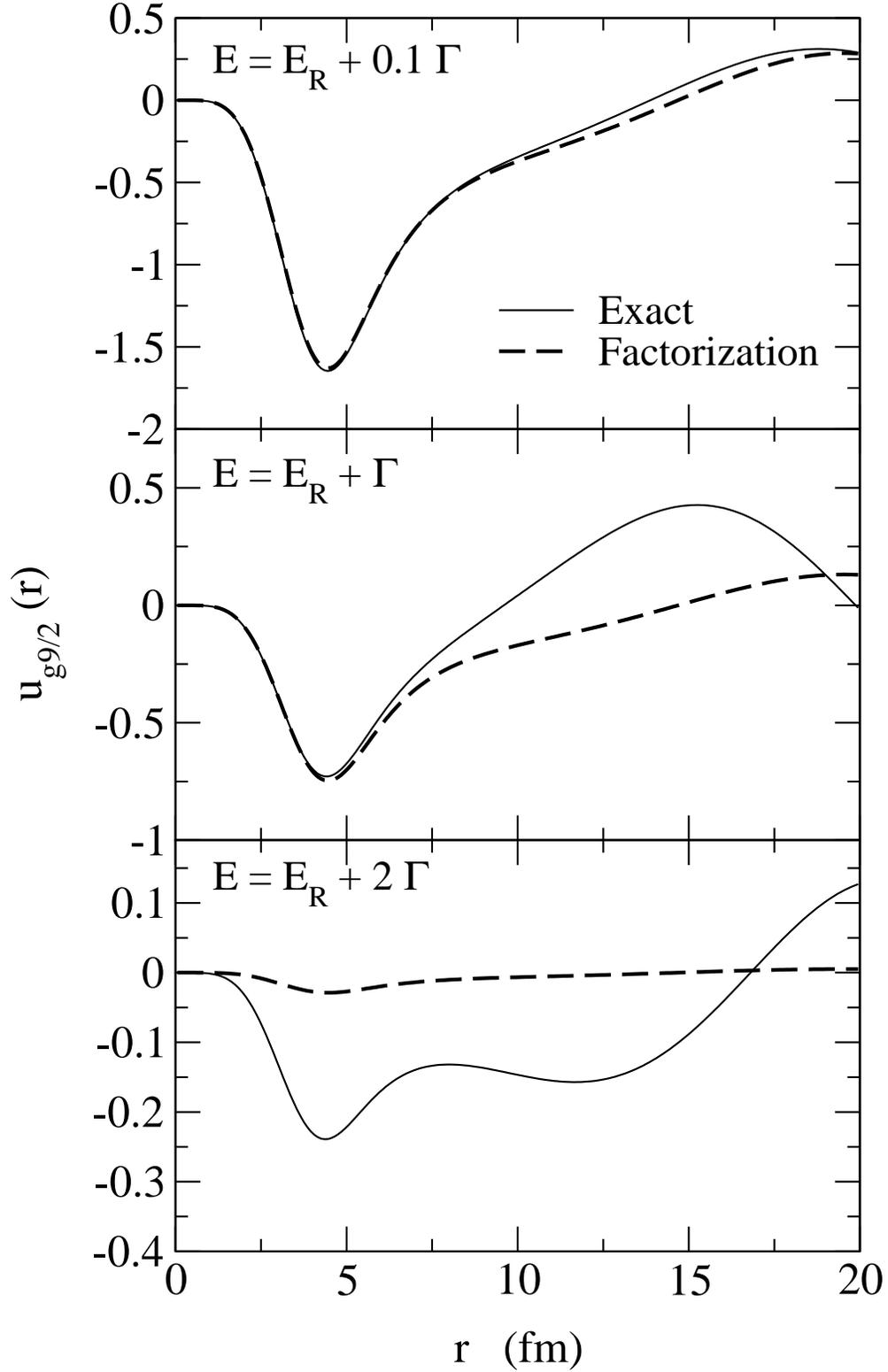}}
  \end{center}
\protect\caption{
The energy dependence of the $g_{9/2}$ component of the eigenchannel 
wave function around the $K^\pi=5/2^+$ 
resonance of $^{44}$S. 
The solid line is the exact wave function, while the dashed line is 
obtained with the factorization formula for a multi-channel wave function.
}
\end{figure}

\end{document}